\documentclass{appolb}
\usepackage{epsfig}
\usepackage{bm}

\newcommand{\bq}{\bm{q}}

\newcommand{\bK}{\bm{K}}
\newcommand{\2}{\frac{1}{2}}

\newcommand{\Eq}{{\,=\,}}
\newcommand{\Kt}{K_\perp}
\begin{document}

\title{Emission angle dependence of HBT radii:\\
Theoretical background and interpretation%
\thanks{Presented at the XXXIII International Symposium on Multiparticle 
Dynamics (ISMD2003), Krak\'ow, Poland, Sept. 5-11, 2003. This work was 
supported by the U.S. Department of Energy under contract DE-FG02-01ER41190.}%
}
\author{Ulrich Heinz
\address{The Ohio State University, Physics Department, 
Columbus, OH 43210, USA}
} 
\maketitle
\begin{abstract}
The Wigner function formalism which relates source size parameters
to experimental ``HBT radii'' extracted from two-particle Bose-Einstein
correlations is generalized to azimuthally deformed and longitudinally 
tilted sources. It is explained how this can be used to complement 
anisotropic flow measurements with relevant space-time information 
on the source. 
\end{abstract}

%%%%%%%%%% Source size porameters and HBT radii %%%%%%%%%%%%%%%%%%%%%%%%%%%%%%
\section{Source size parameters and HBT radii}

\noindent
Two-particle interferometry, which exploits the Bose-Einstein symmetrization 
effects on the production cross section for pairs of identical bosons, has 
become a powerful tool to extract detailed space-time information about 
the freeze-out configuration of the hot and dense 
fireballs formed in relativistic heavy-ion collisions \cite{WH99}. 
Recent exciting data on transverse flow anisotropies in non-central 
heavy-ion collisions and their interpretation as evidence for early 
and efficient thermalization at RHIC energies \cite{KH03} have generated 
new interest in a better understanding of the space-time structure of 
the deformed sources created in these collisions. This requires the 
generalization of the HBT interferometry tool from azimuthally 
symmetric sources (reviewed in \cite{WH99}) to deformed situations. 
Applications of this new formalism to recent RHIC data are
reported in the following article by M.~Lisa.

For chaotic sources (i.e. independent particle emission), the two-particle
Bose-Einstein correlation function $C(\bm{p}_1,\bm{p}_2)=C(\bm{q},\bm{K})$
can be expressed through a Fourier transform of the {\em emission function}
$S(x,\bm{K})$ which describes the single-particle phase-space distribution
at freeze-out \cite{WH99}. $\bK$ is the average momentum of the pair while
$\bq$ is the relative momentum between the two particles. The Fourier 
transform in $\bq$ is restricted by the {\em mass-shell constraint} 
$q^0\Eq{E_1{-}E_2}\Eq\bm{\beta}\cdot\bq$ (where $\bm{\beta}\Eq\bK/K^0$ is
the pair velocity) and therefore not fully invertible. The extraction of
the emission function $S(x,\bm{K})$ from the measured correlation function
$C(\bm{K},\bm{q})$ thus requires additional theoretical input \cite{WH99}.

If the source is dominated by a single length scale (its ``size''), 
the emission function can, for every momentum $\bK$, be approximated
by a Gaussian in space-time whose width parameters form a symmetric
tensor, the {\em spatial correlation tensor} $S_{\mu\nu}(\bK)=
[\langle x_\mu x_\nu\rangle -\langle x_\mu\rangle \langle x_\nu\rangle](\bK)
\equiv\langle\tilde x_\mu\tilde x_\nu\rangle$.
The correlation function is then also a Gaussian in the relative
momentum $\bq$, $C(\bm{q},\bm{K})\Eq1+\exp[-\sum_{i,j=o,s,l}q_iq_j
R_{ij}^2(\bK)]$, where $l$ denotes the beam direction, $o$ the transverse
emission direction $\bK_\perp$ of the pair, and $s$ the third Cartesian
direction perpendicular to $l$ and $o$. The transverse emission direction
$\bK$ is characterized by an azimuthal angle $\Phi$ with respect to
the reaction plane formed by the beam axis and the impact parameter $\bm{b}$.
In collisions between spherical nuclei the emission function is reflection 
symmetric with respect to the reaction plane. This and other, more
specific symmetries of the emission function are most easily expressed
in a reaction-plane-fixed coordinate system where $z$ points along the
beam direction, $x$ along the impact parameter, and $y$ perpendicular
to the reaction plane. On the other hand, the correlation radii
$R_{ij}^2(\bK)$ are more easily interpreted in the $(o,s,l)$ system
because the $q_s$-dependence, being transverse to the pair velocity
$\bm{\beta}$, is not affected by the mixing of space and time induced
by the mass-shell constraint into the Fourier transform. The $(o,s,l)$ 
and $(x,y,z)$ systems are rotated with respect to each other by the 
azimuthal emission angle $\Phi$.

The Fourier transform between the emission and correlation functions
the leads to the following relations between the ``HBT radii'' 
$R_{ij}^2(\bK)$ and the components of the spatial correlation tensor 
$S_{\mu\nu}(\bK)$ \cite{W97}:
  \begin{eqnarray}
    R_s^2 &=& \textstyle{\2}(S_{xx}{+}S_{yy}) 
            - \textstyle{\2}(S_{xx}{-}S_{yy})\cos(2\Phi)
            - S_{xy} \sin(2\Phi)
  \nonumber\\
    R_o^2 &=& \textstyle{\2}(S_{xx}{+}S_{yy}) 
            + \textstyle{\2}(S_{xx}{-}S_{yy})\cos(2\Phi)
            + S_{xy} \sin(2\Phi)
  \nonumber\\
          &&- 2\beta_\perp (S_{tx} \cos\Phi{+}S_{ty} \sin\Phi)
             + \beta_\perp^2 S_{tt}, 
  \nonumber\\
     R_{os}^2 &=& S_{xy} \cos(2\Phi) 
        - \textstyle{\2} \left(S_{xx}{-}S_{yy}\right)\sin(2\Phi)
        + \beta_\perp (S_{tx} \sin\Phi{-}S_{ty} \cos\Phi), 
  \nonumber\\
    R_{l}^2 &=& S_{zz} -2 \beta_l S_{tz} + \beta_l^2 S_{tt}, 
  \nonumber\\
    R_{ol}^2 &=& \left( S_{xz}{-}\beta_l S_{tx}\right) \cos\Phi
               + \left( S_{yz}{-}\beta_l S_{ty}\right) \sin\Phi
               - \beta_\perp S_{tz} 
               + \beta_l\beta_\perp S_{tt},
  \nonumber\\
    R_{sl}^2 &=& \left(S_{yz}{-}\beta_l S_{ty}\right) \cos\Phi
               - \left(S_{xz}{-}\beta_l S_{tx}\right) \sin\Phi .
  \end{eqnarray}
These relations display the {\em explicit} $\Phi$-dependence arising
from the mentioned rotation between the $(x,y,z)$ and $(o,s,l)$ systems,
but hide the {\em implicit} $\Phi$-dependence of the spatial correlation 
tensor components $S_{\mu\nu}(\bK)=S_{\mu\nu}(Y,\Kt,\Phi)$. The total emission
angle dependence of the HBT radii results from a combination of both
explicit and implicit $\Phi$-dependences \cite{W97,HHLW02}.

The implicit $\Phi$-dependence of the spatial correlation tensor is
restricted by symmetries of the source \cite{HHLW02}. It is a 
relativistic effect associated with an azimuthal spatial source 
deformation superimposed by strong transverse collective flow
\cite{W97,H02} which vanishes with the $4^{\rm th}$ power of the 
transverse flow velocity $v_T/c$ for weak or no collective expansion 
\cite{H02,LHW00}. 

%%%%%%%%%% Azimuthal oscillations and harmonic analysis %%%%%%%%%%%%%%%%%%%%%%
\section{Azimuthal oscillations and harmonic analysis}

\noindent
A full analysis of the symmetry constraints on $S_{\mu\nu}(Y,\Kt,\Phi)$
for symmetric collisions between spherical nuclei and for pairs detected
in a symmetric rapidity window around $Y\Eq0$ can be found in 
Ref.~\cite{HHLW02}. One finds the following {\em most general} form
for the azimuthal oscillations of the HBT radii:
\begin{eqnarray}
 \label{24}
   R_s^2 &\!\!=\!\!& R_{s,0}^2 + {\textstyle2\sum_{n=2,4,6,\dots}} 
   R_{s,n}^2\cos(n\Phi),
   \quad
   R_{os}^2 = {\textstyle2\sum_{n=2,4,6,\dots}} R_{os,n}^2\sin(n\Phi),
 \nonumber\\
   R_o^2 &\!\!=\!\!& R_{o,0}^2 + {\textstyle2\sum_{n=2,4,6,\dots}} 
   R_{o,n}^2\cos(n\Phi),
   \quad
   R_{ol}^2 = {\textstyle2\sum_{n=1,3,5,\dots}} R_{ol,n}^2\cos(n\Phi),
 \nonumber\\
   R_l^2 &\!\!=\!\!& R_{l,0}^2 + {\textstyle2\sum_{n=2,4,6,\dots}}
   R_{l,n}^2\cos(n\Phi),
   \quad
   R_{sl}^2 = {\textstyle2\sum_{n=1,3,5,\dots}} R_{sl,n}^2\sin(n\Phi).\ 
\end{eqnarray}
We see that only even {\em or} odd sine {\em or} cosine terms occur, but
no mixtures of such terms. Statistical errors in the resolution of the 
reaction plane angle as well as finite angular bin sizes in $\Phi$ tend 
to reduce the actually measured oscillation amplitudes; fortunately, 
these dilution effects can be fully corrected by a model-independent 
correction algorithm \cite{HHLW02}.
A Gaussian fit to the thus corrected correlation function, binned in $Y$,
$\Kt$ and emission angle $\Phi$, then yields the ``true'' HBT radius
parameters $R_{ij}^2(Y,\Kt,\Phi)$ from which the $n^{\rm th}$ order 
azimuthal oscillation amplitudes are extracted via
%\begin{equation}
$
  R_{ij,n}^2(Y,\Kt) = \frac{1}{n_{\rm bin}} \sum_{j=1}^{n_{\rm bin}} 
          R_{ij}^2(Y,\Kt,\Phi_j) {\rm osc}(n \Phi_j).
$
%\end{equation}
Here $n_{\rm bin}$ indicates the number of (equally spaced) $\Phi$ bins 
in the data and ${\rm osc}(n\Phi_j)$ stands for $\sin(n\Phi_j)$ or 
$\cos(n\Phi_j)$ as appropriate, see Eqs.\,(\ref{24}). (Note that 
Nyquist's theorem limits the number of harmonics that can be extracted 
to $n\leq n_{\rm bin}$.) 

%%%%%%%%%%  HBT oscillation amplitudes and source shape %%%%%%%%%%%%%%%%%%%%%%
\section{HBT oscillation amplitudes and source shape}

\noindent
We would like to relate the azimuthal oscillation amplitudes of the 
6 HBT radius parameters to the geometric and dynamical anisotropies 
of the source, as reflected in the azimuthal oscillations of the 10 
independent components of the spatial correlation tensor. Their allowed
oscillation patterns at midrapidity $Y\Eq0$ are given by \cite{HHLW02}
\begin{eqnarray}
\label{A-J}
  A(\Phi) \equiv \textstyle{\2}\langle\tilde x^2{+}\tilde y^2\rangle
  &=& \textstyle{A_0{+}2\sum_{n\geq2,{\rm even}} A_n\cos(n\Phi)},
\nonumber\\   
  B(\Phi) \equiv \textstyle{\2}\langle\tilde x^2{-}\tilde y^2\rangle
  &=& \textstyle{B_0{+}2\sum_{n\geq2,{\rm even}} B_n\cos(n\Phi)},
\nonumber\\  
 C(\Phi) \equiv \langle\tilde x\tilde y\rangle
  &=& \textstyle{\phantom{A_0{+}}2\sum_{n\geq2,{\rm even}} C_n\sin(n\Phi)},
\nonumber\\  
  D(\Phi) \equiv \langle\tilde t^2\rangle
  &=& \textstyle{D_0{+}2\sum_{n\geq2,{\rm even}} D_n\cos(n\Phi)},
\nonumber\\  
  E(\Phi) \equiv \langle\tilde t\tilde x\rangle
  &=& \textstyle{\phantom{A_0{+}}2\sum_{n\geq1,{\rm odd}} E_n\cos(n\Phi)},
\nonumber\\  
  F(\Phi) \equiv \langle\tilde t\tilde y\rangle
  &=& \textstyle{\phantom{A_0{+}}2\sum_{n\geq1,{\rm odd}} F_n\sin(n\Phi)},
\nonumber\\  
  G(\Phi) \equiv \langle\tilde t\tilde z\rangle
  &=& \textstyle{\phantom{A_0{+}}2\sum_{n\geq1,{\rm odd}} G_n\cos(n\Phi)},
\nonumber\\  
  H(\Phi) \equiv \langle\tilde x\tilde z\rangle
  &=& \textstyle{H_0{+}2\sum_{n\geq2,{\rm even}}\!H_n\cos(n\Phi)},
\nonumber\\  
  I(\Phi) \equiv \langle\tilde y\tilde z\rangle
  &=& \textstyle{\phantom{A_0{+}}2\sum_{n\geq2,{\rm even}} I_n\cos(n\Phi)},
\nonumber\\  
  J(\Phi) \equiv \langle\tilde z^2\rangle
  &=& \textstyle{J_0\,{+}2\sum_{n\geq2,{\rm even}} J_n\cos(n\Phi)}.
\end{eqnarray}
The missing terms in the sums over $n$ have amplitudes which are odd
functions of $Y$ and vanish at midrapidity. They do, however, contribute 
to the HBT radii if the data are averaged over a finite, symmetric 
rapidity window around $Y\Eq0$ \cite{HHLW02}. Their contributions can 
be eliminated by varying the width $\Delta Y$ of this rapidity window 
and extrapolating quadratically to $\Delta Y\to0$. Note that 
$C_0\Eq{E}_0\Eq{F}_0\Eq{G}_0\Eq{I_0}\Eq0$ by symmetry, i.e. the 
corresponding components of $S_{\mu\nu}$ oscillate around zero.

The oscillation amplitudes of the HBT radii relate to the oscillation
amplitudes of the source parameters as follows: For the odd harmonics
$n\Eq1,3,5,\dots$ we have
\begin{eqnarray}
  R_{ol,n}^2 &=& {\textstyle\2}
 \langle H_{n{-}1}{+}H_{n{+}1}{-}I_{n{-}1}{+}I_{n{+}1}
 -\beta_l(E_{n{-}1}{+}E_{n{+}1}{-}F_{n{-}1}{+}F_{n{+}1})\rangle
\nonumber\\
  && -\langle  \beta_\perp G_n -\beta_l D_n\rangle,
\\\nonumber
  R_{sl,n}^2 &=& {\textstyle\2}
 \langle{-}H_{n{-}1}{+}H_{n{+}1}{+}I_{n{-}1}{+}I_{n{+}1}
 -\beta_l({-}E_{n{-}1}{+}E_{n{+}1}{+}F_{n{-}1}{+}F_{n{+}1})\rangle,
\end{eqnarray}
whereas the even harmonics $n=0,2,4,\dots$ satisfy
\begin{eqnarray}
  R_{s,n}^2 &=& \langle A_n\rangle 
  +{\textstyle\2}\langle {-}B_{n{-}2}{-}B_{n{+}2}{+}C_{n{-}2}{-}C_{n{+}2}
  \rangle,
\nonumber\\
  R_{o,n}^2 &=& \langle A_n\rangle 
 +{\textstyle\2} \langle B_{n{-}2}{+}B_{n{+}2}{-}C_{n{-}2}{+}C_{n{+}2}\rangle
\nonumber\\
 && - \beta_\perp\langle E_{n{-}1}{+}E_{n{+}1}{-}F_{n{-}1}{+}F_{n{+}1}\rangle
                + \beta_\perp^2 \langle D_n\rangle,
\nonumber\\
  R_{os,n}^2 &=& {\textstyle\2}\langle{-}B_{n{-}2}{+}B_{n{+}2}
   {+}C_{n{-}2}{+}C_{n{+}2}
  + \beta_\perp(E_{n{-}1}{-}E_{n{+}1}{-}F_{n{-}1}{-}F_{n{+}1})\rangle,
\nonumber\\
  R_{l,n}^2 &=& \langle J_n\rangle -2\langle\beta_l G_n\rangle
   +\langle\beta_l^2 D_n\rangle.
\end{eqnarray}
In these relations it is understood that all negative harmonic coefficients
$n<0$ as well as $C_0,E_0,F_0,G_0$ and $I_0$ are zero. The angular brackets 
$\langle\dots\rangle$ indicate an average over a finite, symmetric rapidity 
window around $Y\Eq0$. The terms involving the longitudinal pair velocity 
$\beta_l$ vanish quadratically as the width $\Delta Y$ of that window 
shrinks to zero. 

Even after extrapolating to $Y\Eq0$ in this way, we have still many
more source parameters than measurable HBT amplitudes. One counts easily 
that up to $n\Eq2$ there are 9 measurable Fourier coefficients which 
(at $Y\Eq0$) depend on 19 source amplitudes. From there on, increasing 
$n$ by 2 yields 6 additional measured amplitudes which depend on 10 
additional source amplitudes. This lack of analysis power is an 
intrinsic weakness of the HBT microscope and due to the fundamental
restrictions arising from the mass-shell constraint 
$q^0\Eq\bm{\beta}\cdot\bq$. The reconstruction of the source
thus must necessarily rely on additional assumptions. 

One such assumption which may not be too unreasonable is that the 
emission duration $D\Eq\langle\tilde t^2\rangle$ is approximately
independent of emission angle and that the source is sufficiently
smooth that higher order harmonics $n\geq3$ of $S_{\mu\nu}$ can be
neglected. Such source properties would result in the ``Wiedemann sum
rule'' \cite{W97} 
\begin{equation}
  R_{o,2}^2-R_{s,2}^2 + 2 R_{os,2}^2 = 0
\end{equation}
which can be experimentally tested. If verified for all $\Kt$ it would
provide strong support for the underlying assumptions on the source. 
In this case we can measure 3 azimuthally averaged HBT radii and
5 independent oscillation amplitudes with $n\leq2$, depending
on 14 source parameters of which 5 can be eliminated by going to 
$\Kt\Eq\beta_\perp\Eq0$ (see \cite{HHLW02} for explicit expressions). 
This makes the geometry of the effective source for particles with 
$\Kt\Eq0$ ``almost solvable'' \cite{fn1}, as confirmed by hydrodynamical 
calculations \cite{HK02} which show that at $\Kt\Eq0$ the 
effective emission region closely tracks the overall geometry of 
the source even if it is strongly and 
anisotropically expanding. 

The source geometry can be completely reconstructed from HBT data if 
transverse flow is so weak that all implicit $\Phi$-dependence (i.e. 
all higher harmonics $n\geq1$) of $S_{\mu\nu}$ can be neglected. In 
this case one obtains at $Y\Eq0$ the ``geometric relations'' \cite{LHW00} 
\begin{eqnarray}
  &&R_{s,0}^2 = A_0 = \textstyle{\2}\langle\tilde x^2{+}\tilde y^2\rangle_0,
\quad\ 
  R_{o,0}^2-R_{s,0}^2 = \beta_\perp^2 D_0 
                      = \beta_\perp^2 \langle\tilde t^2\rangle_0,
\nonumber\\
  &&R_{l,0}^2 = J_0 = \langle\tilde z^2\rangle_0,
\qquad\qquad 
    R_{ol,1}^2 = - R_{sl,1}^2 = \textstyle{\2} H_0
                  = \textstyle{\2}\langle\tilde x\tilde z\rangle_0,
\nonumber\\
  &&R_{o,2}^2 = - R_{s,2}^2 = - R_{os,2}^2 = \textstyle{\2} B_0
    = \textstyle{\frac{1}{4}}\langle\tilde x^2{-}\tilde y^2\rangle_0.
\end{eqnarray}
$A_0$ describes the average transverse size and $B_0$ (which generates 
a second-order harmic in the transverse HBT radii) the transverse 
deformation of the source. $H_0$ generates a first-order harmonic 
in the $ol$ and $sl$ cross terms and describes a longitudinal tilt of 
the source away from the beam direction \cite{LHW00}. Such a tilt was 
found in Au+Au collisions at the AGS \cite{E895}. Its sign yielded 
important information on the kinetic pion production mechanism 
\cite{LHW00,E895}.

%%%%%%%%%%%%%%%%%%%%%%%%%%%%%%%%% Conclusions %%%%%%%%%%%%%%%%%%%%%%%%%%%%%%%
\section{Conclusions}

Azimuthally sensitive HBT interfrometry is a powerful tool for
analyzing the dynamic origin and space-time manifestations of 
the strong anisotropic collective flow seen in single-particle 
spectra at RHIC. Symmetries strongly constrain the azimuthal Fourier 
series of the emission angle dependent HBT radii. This is helpful 
but not sufficient for a fully model independent reconstruction of 
the source deformations from such data. To separate temporal from geometric
contributions to the HBT radii, the source must satisfy special properties.
One can test for them experimentally, albeit not in a fully
model-independent way. Hydrodynamic simulations \cite{HK02} (which were
reported at the meeting but are not included here) show that at RHIC and 
LHC energies emission from the source at non-zero
transverse momentum is very strongly surface dominated (``source opacity'').
To obtain an unambiguous estimate of the transverse deformation of the source
at freeze-out one should study the $\Phi$-oscillations of the HBT radii
at very small $\Kt$. A possible longitudinal tilt of the source away from
the beam direction manifests itself through first-order harmonics
in $R_{ol}^2$ and $R_{ol}^2$. The $\Kt$-dependence of the oscillation 
patterns of the transverse HBT radius parameters can be analyzed to
obtain evidence for or against a faster expansion of the source in-plane
than out-of-plane due to anisotropic collective flow \cite{HK02}.   

%%%%%%%%%%%%%%%%%%%%%%%% References %%%%%%%%%%%%%%%%%%%%%%%%%%%%%%%%%%%%%%%%%%%


\begin{thebibliography}{99}

\bibitem{WH99}
  U.A.~Wiedemann and U.~Heinz,
  %``Particle interferometry for relativistic heavy-ion collisions,''
  Phys.\ Rept.\  {\bf 319}, 145 (1999);\\
  %%CITATION = NUCL-TH 9901094;%%
  U.~Heinz and B.V.~Jacak,
  %``Two-particle correlations in relativistic heavy-ion collisions,''
  Ann.\ Rev.\ Nucl.\ Part.\ Sci.\  {\bf 49}, 529 (1999);\\
  %%CITATION = NUCL-TH 9902020;%%
  B.~Tom\'a\v sik and U.A. Wiedemann,
  hep-ph/0210250.
  %%CITATION = HEP-PH 0210250;%%

\bibitem{KH03}
  P.~F.~Kolb and U.~Heinz,
%``Hydrodynamic description of ultrarelativistic heavy-ion collisions,''
  arXiv:nucl-th/0305084.
%%CITATION = NUCL-TH 0305084;%%

\bibitem{W97}
  U.A.~Wiedemann,
  %``Two-particle interferometry for noncentral heavy-ion collisions,''
  Phys.\ Rev.\ C {\bf 57}, 266 (1998).
  %%CITATION = NUCL-TH 9707046;%%

\bibitem{HHLW02}
  U. Heinz, A. Hummel, M.A. Lisa and U.A. Wiedemann, Phys. Rev. C {\bf 66},
  044903 (2002).
  %%CITATION = NUCL-TH 0207003;%%

\bibitem{H02}
  A. Hummel, M.Sc. Thesis, Ohio State University, 2002, unpublished.

\bibitem{LHW00}
  M.A.~Lisa, U.~Heinz and U.A.~Wiedemann,
  %``Tilted pion sources from azimuthally sensitive HBT interferometry,''
  Phys.\ Lett.\ B {\bf 489}, 287 (2000).
  %%CITATION = NUCL-TH 0003022;%%

\bibitem{fn1}
  Note that the limit $\Kt\to0$ eliminates all influence from the 
  temporal structure of the source; the emission duration and 
  correlations between position and time at freeze-out
  must be extracted from correlation data at non-zero $\Kt$ where,
  however, the explicit $\beta_\perp$-dependence associated with 
  factors $t$ in the variances $\langle\tilde x_\mu\tilde x_\nu\rangle$
  interferes with the implicit $\Kt$-dependence in the spatial
  variances which result from collective expansion flow \cite{WH99}.
  Models for disentangling these different contributions to the 
  $\Kt$-dependence have been extensively studied for azimuthally
  symmetric sources \cite{WH99}, but must be generalized for
  azimuthally deformed sources \cite{LR03}.

\bibitem{LR03}
  M.A. Lisa and F. Reti\'ere, to be published.

\bibitem{HK02}
   U.~Heinz and P.F.~Kolb,
%``Emission angle dependent pion interferometry at RHIC and beyond,''
   Phys.\ Lett.\ B {\bf 542}, 216 (2002).
%%CITATION = HEP-PH 0206278;%%

\bibitem{E895}
  M.A.~Lisa {\it et al.}  [E895 Collaboration],
  %``Azimuthal dependence of pion interferometry at the AGS,''
  Phys.\ Lett.\ B {\bf 496}, 1 (2000).
  %%CITATION = NUCL-EX 0007022;%%

\end{thebibliography}
\end{document}